\documentclass[%
 reprint,
nofootinbib,
 amsmath,amssymb,
 aps,
]{revtex4-2}

\usepackage{graphicx}
\usepackage{dcolumn}
\usepackage{bm}

\usepackage{hyperref}
\usepackage{comment}
\usepackage[dvipsnames]{xcolor}
\usepackage{orcidlink}
\usepackage{slashed}
\usepackage{soul}
\hypersetup{
  colorlinks=true,
  linkcolor=red,
  citecolor=blue,
  urlcolor=blue,
 hypertexnames=false 
}

\begin{document}


\hfill TTP26-003,  P3H-26-007

\title{Neutrino Masses with Enhanced $B-L$ Symmetry}

\author{Xiyuan Gao\,\orcidlink{0000-0002-1361-4736}, and Amir N. Khan\,\orcidlink{0000-0003-1039-3349}}
\email{xiyuan.gao@kit.edu,\\ amir.khan@kit.edu}
\affiliation{\normalsize \it 
 Institut f\"ur Theoretische Teilchenphysik (TTP),
  Karlsruher Institut f\"ur Technologie (KIT), 76131 Karlsruhe, Germany.}



\begin{abstract}
Assuming all three known neutrinos are Dirac fermions, $U(1)_{B-L}$ can be an exact symmetry. We show that, if the condition of charge quantization is relaxed, the anomaly-free $B-L$ charges of two out of three right-handed neutrinos can be enhanced by arbitrarily large factors, while all other fermions retain their canonical charges. 
We call this setup as `enhanced $B-L$ symmetry' and promote it to be local.
As long as this enhanced $B-L$ gauge symmetry remains unbroken, neutrinos stay chiral and massless at low energies. Nonzero neutrino masses then require sub-eV-scale symmetry breaking order parameters, which we associate with gravity-induced neutrino condensate.
If the enhancement is large and the $B-L$ gauge boson $A'$ is lighter than the heaviest neutrino, then the neutrino decay into $A'$ directly constrains the gauge coupling, which can be significantly stronger than the baryon-based fifth-force tests. Through kinetic mixing with the photon, $A'$ can also mediate neutrino--electron and coherent neutrino--nucleus scatterings, leading to possible signatures in neutrino observatories and dark matter detectors. 
\end{abstract}

\maketitle


\noindent

\textit{Introduction:}~The origin of neutrino masses is an open question in particle physics. 
The observation of neutrino oscillations~\cite{Super-Kamiokande:1998kpq, SNO:2001kpb, SNO:2002tuh} implies that the neutrinos are massive and seven to thirteen orders of magnitude lighter than the charged fermions.
On the other hand, no long-range interactions except for gravity and electromagnetism have been observed in nature~\cite{ParticleDataGroup:2024cfk}, despite the fact that no fundamental principle forbids them. 
Assuming all three active neutrinos are Dirac type particles, the $U(1)_{B-L}$, where $B$ and $L$ respectively denote the baryon and lepton numbers, can be an exact symmetry of nature~\cite{Babu:1989tq, Foot:1992ui}.
If this symmetry is gauged and remains unbroken down to sufficiently low energies, the local $U(1)_{B-L}$ symmetry then manifests itself as an additional long-range force, often called `the fifth force'.


An extensively studied framework that naturally accounts for the smallness of neutrino masses and the associated \(U(1)_{B-L}\) structure is the seesaw mechanism~\cite{Minkowski:1977sc, Yanagida:1979as, Gell-Mann:1979vob, Mohapatra:1979ia}, which admits embeddings in left--right symmetric models~\cite{Mohapatra:1979ia} as well as in SO(10) grand unified theories~\cite{Witten:1979nr}. In its minimal realization, the three right-handed neutrinos $\nu_R$, singlets under the $SU(3)_c \times SU(2)_L \times U(1)_Y$ gauge group and charged under \(B-L\), acquire large Majorana masses and decouple from low-energy dynamics. Consequently, lepton number is violated and the left-handed neutrinos $\nu_L$ acquire tiny Majorana masses through the dimension-5 Weinberg operator~\cite{Weinberg:1979sa}, leaving only the $U(1)$ symmetry of electromagnetism exact~\cite{Babu:1989tq, Foot:1992ui}.


While the seesaw paradigm is conceptually appealing, the non-observation of a fifth force does not exclude the existence of additional gaugeable $U(1)$ symmetries. 
Rather, the associated interactions may be extremely weak and thus evade current experimental sensitivity, while still leaving imprints in the form of violations of the weak equivalence principle~\cite{Schlamminger:2007ht, ADELBERGER2009102, MICROSCOPE:2022doy} and/or deviations from the inverse-square law of gravity~\cite{Kapner:2006si, Sushkov:2011md, Yang:2012zzb, Chen:2014oda, Tan:2020vpf}.\footnote{If the long-range force is lepton-flavor non-universal, neutrino oscillation experiments can also provide stringent constraints; see, e.g.,~\cite{Joshipura:2003jh, Grifols:2003gy, Gonzalez-Garcia:2006vic, Bandyopadhyay:2006uh, Heeck:2010pg, Davoudiasl:2011sz, Wise:2018rnb, Bustamante:2018mzu, Dror:2020fbh, Coloma:2020gfv, Agarwalla:2023sng, ESSnuSB:2025shd, Garg:2026gwx}.}
On the other hand, after electroweak symmetry breaking, neutrino masses need not originate from bare mass terms that explicitly break the chiral symmetry of otherwise massless neutrinos. 
Instead, effective neutrino mass terms can be generated through non-perturbative gravitational effects~\cite{Dvali:2016uhn, Dvali:2017mpy} or scattering with background dark matter particles~\cite{Davoudiasl:2018hjw, Ge:2019tdi, Choi:2019zxy, Choi:2020ydp}. Such mechanisms naturally predict time-varying neutrino masses~\cite{Lorenz:2018fzb, Sen:2023uga} in light of the latest combined results from the DESI~\cite{DESI:2024mwx} and Planck~\cite{Planck:2018vyg} collaborations, which impose increasingly stringent upper bounds on the sum of neutrino masses.

The possibility of effective neutrino mass motives us to reconsider the assignment of the $U(1)_{B-L}$ charges, denoted by $Q_{B-L}$. 
Conventionally, it is assumed that $Q_{B-L}$ of the right-handed neutrinos $\nu_R$ are identical to that of the left-handed lepton doublet $\ell_L$, so that the Dirac neutrino mass term $m_D\overline{\ell}_L\nu_R$ can exist. Once this requirement is relaxed, the anomaly-free $Q_{B-L}$ assignment for $\nu_R$ becomes less constrained; see, e.g.~\cite{Montero:2007cd, Montero:2011jk, Ma:2014qra, Allanach:2018vjg}.
In this work, we find that the $Q_{B-L}$ values for two generations of $\nu_R$ can be enhanced by arbitrarily large factors, while those for quarks and charged leptons remain at their canonical values. 
To our knowledge, this result represents a previously unexplored regime.
The $B-L$ gauge interaction, if realized in nature, can therefore be sizable for neutrinos while remaining extremely feeble for baryons and charged leptons. We argue that the $B-L$ gauge coupling strength, whether enhanced or not, should be interpreted as a fundamental parameter of nature, since no fundamental principle forbids gauging this exact symmetry. If enhanced, its upper bound---possibly as large as $\mathcal{O}(1)$---remains largely unexplored, even in the presence of existing phenomenological constraints.

We find that the experimental implications of this enhanced interaction distinct from those of a broad class of neutral gauge boson models discussed in the literature, see, e.g., Refs.~\cite{He:1990pn, He:1991qd, Foot:1994vd, Ma:1997nq, Araki:2012ip, Heeck:2014zfa,Batell:2016zod, Farzan:2016wym, Ballett:2019xoj, Herbermann:2025uqz}. In our framework, neutrino decay and neutrino--electron ($\nu$--$e$) elastic scattering as well as coherent elastic neutrino--nucleus scattering (CE$\nu$NS), provide the most stringent constraints.
Consequently, currently operating and next-generation neutrino observatories and telescopes—such as DUNE~\cite{DUNE:2020lwj, DUNE:2015lol, LBNE:2013dhi, Ghoshal:2020hyo}, JUNO~\cite{JUNO:2015zny, JUNO:2021vlw, Abrahao:2015rba}, Hyper-Kamiokande~\cite{Hyper-KamiokandeProto-:2015xww}, IceCube-Gen2~\cite{IceCube-Gen2:2020qha}, and ESSnuSB~\cite{Chakraborty:2020cfu}—as well as dark matter direct-detection experiments, including liquid noble-gas detectors such as PandaX-4T~\cite{PandaX:2024qfu, PandaX:2024cic}, XENONnT~\cite{XENON:2022ltv, XENON:2023cxc}, LUX-ZEPLIN (LZ)~\cite{LZ:2025igz, LZ:2024zvo}, DarkSide-50~\cite{DarkSide-50:2022qzh}, and DARWIN~\cite{DARWIN:2020bnc}, as well as cryogenic solid-state detectors such as SENSEI~\cite{SENSEI:2023zdf} and DAMIC-M~\cite{DAMIC-M:2025luv, DAMIC-M:2025ltz}, are especially important to test our framework.

\textit{Enhanced $B-L$ Symmetry}~If the three generation of neutrinos are Dirac type, the exact symmetries of nature can be $SU(3)_{c}\times U(1)_{\text{QED}}\times U(1)_{B-L}$. 
By definition, all quarks carry baryon number $B=1/3$, while all charged leptons carry lepton number $L=1$. As $SU(2)_L$ partners of the charged leptons, the left-handed neutrinos $\nu_L$ inherit lepton number $L=1$. The charge assignments are shown in Table~\ref{charge1}. 
\begin{table}[t!]
    \centering
\renewcommand\arraystretch{2}
    \begin{tabular}{c | c c c  }
    \hline
    & $~~Q_{B-L}~~$  & $~~Q_{B-L}'~~$ & $Q_{B-L}''$ \\
    \hline
    $(Q_L^{\alpha i}, u_R^{\alpha i}, d_R^{\alpha i})$  & $+\frac{1}{3}$ &    $+\frac{1}{3}$ & $+\frac{1}{3}$ \\
    $(e_L, \nu_{eL}, e_R)  $ & $-1$  & $-1$ & $-1$ \\
    $(\mu_L, \nu_{\mu L}, \mu_R)  $  & $-1$ & $-1$ & $-1$  \\
    $(\tau_L, \nu_{\tau L}, \tau_R)  $ & $-1$ &   $-1$ & $-1$ \\
    $\nu_{eR}  $ & $-1$  &   $+5$ & $-3+\mathcal{O}(\epsilon^2)$  \\
    $ \nu_{\mu R}  $ & $-1$ &   $-4$ & $+\frac{1}{\epsilon}+\mathcal{O}(\epsilon)$  \\
    $\nu_{\tau R}  $ & $-1$ &    $-4$ & $-\frac{1}{\epsilon}+\mathcal{O}(\epsilon)$ \\
   \hline
\end{tabular}
    \caption{Certain $B-L$ charge assignments to the quarks, charged leptons, and neutrinos, where \(\alpha = (r,b,g)\) is the color index and \(i = (1,2,3)\) is the flavor index.}
    \label{charge1}
\end{table}

The $B-L$ symmetry is anomaly-free provided that the $B-L$ charge for the three generations of $\nu_R$ satisfy~\cite{Montero:2007cd}:
\begin{equation}
\label{neutrinocharge}
\begin{aligned}
    Q_{\nu_{eR}}+Q_{\nu_{\mu R}}+Q_{\nu_{\tau R}}~&=~-3, \\
    Q_{\nu_{eR}}^3+Q_{\nu_{\mu R}}^3+Q_{\nu_{\tau R}}^3~&=~-3.\\     
\end{aligned}
\end{equation}
Here, we use $(e,\mu, \tau)$ only to label the three generations of $\nu_R$. These can not be distinguished by the electroweak interactions.
The conventional solution to these conditions is $Q_{\nu_{eR}}=Q_{\nu_{\mu R}}=Q_{\nu_{\tau R}}=-1$, as shown the 1st column of Table~\ref{charge1}.  
In this case, the $B-L$ symmetry is vectorial, allowing Dirac neutrino mass terms while forbidding the Majorana mass terms. 
Ref.~\cite{Montero:2007cd} first pointed out an alternative solution to Eq~(\ref{neutrinocharge}) in the integer domain. Up to permutations, the $B-L$ charges for the three $\nu_R$ fields can take values $\left(Q_{\nu_{eR}}, Q_{\nu_{\mu R}}, Q_{\nu_{\tau R}}\right)=(+5, -4,-4)$, as shown in the 2nd column of Table~\ref{charge1}. This setup is chiral for the neutrinos, under which left-handed and right-handed neutrinos have different charges, and forbids both Dirac and Majorana neutrino mass terms.

The $B-L$ charges are not necessarily quantized and in general can take real values. Eq~(\ref{neutrinocharge}) admits infinitely many solutions . 
We identify a novel solution which, to our knowledge, has not been explored previously.
In this solution, we take $Q_{\nu_{e R}}\to -3$ and  $Q_{\nu_{\mu R}}\to -Q_{\nu_{\tau R}}$, and find the following resulting charges: 
\begin{equation}
    \label{solReal}
    \begin{aligned}
        Q_{\nu_{eR}}~&=~-3-8\epsilon^2, \\
        Q_{\nu_{\mu R}}~&=~ +\frac{\sqrt{\epsilon ^2+1} }{\epsilon }\left(4 \epsilon ^2+1\right)+4 \epsilon^2 \\
            ~&\approx~+\left(\frac{1}{\epsilon}+\frac{9}{2}\epsilon+4\epsilon^2+\mathcal{O}(\epsilon^3)\right), \\
        Q_{\nu_{\tau R}}~&=~-\frac{\sqrt{\epsilon ^2+1} }{\epsilon }\left(4 \epsilon ^2+1\right)+4 \epsilon ^2\\
        ~&\approx~-\left(\frac{1}{\epsilon}+\frac{9}{2}\epsilon-4\epsilon^2+\mathcal{O}(\epsilon^3)\right), \\  
    \end{aligned}
\end{equation}
also satisfies Eq.~\ref{neutrinocharge}. In the 3rd column of Table~\ref{neutrinocharge}, we show this charge set-up in the limit $\epsilon\to0$. 
We note that assigning $Q_{\nu_{e R}}\to -3$ is equivalent to $Q_{\nu_{\mu R}}\to -3$ or $Q_{\nu_{\tau R}}\to -3$, because the $B-L$ interaction is flavor universal for the other particles listed in Table~\ref{neutrinocharge}.
Importantly, $Q_{B-L}$ for any two of the three generations of $\nu_R$ can be enhanced by arbitrarily large factors. 
We refer to this this setup as the `enhanced $B-L$ symmetry' and promote it to a gauge symmetry.

Note that $Q_{B-L}\gg1$ does not imply a violation of perturbative unitarity, because the $1/\epsilon$ factor can be absorbed by defining an effective coupling $g_{\text{eff}}^{\nu }\equiv\epsilon^{-1}g_{B-L}$.
Then, if $\epsilon=0$, the enhanced $B-L$ symmetry reduces to the $U(1)_{L_{\mu}-L_{\tau}}$ symmetry under which only two generations of $\nu_R$ fields carry opposite charges. 
This $L_{\mu}-L_{\tau}$ symmetry is non-chiral and therefore automatically anomaly-free, providing a consistency check for our solution. 
The enhanced $B-L$ symmetry is distinct from the $L_\mu - L_\tau$ symmetry for $\nu_R$: for $\epsilon = 0$, both Majorana and Dirac neutrino mass terms are allowed, whereas for $0 < \epsilon \ll 1$, these mass terms are forbidden.

We emphasize that $\epsilon$ should not be regarded as a conventional free parameter, such as a mixing angle. Rather, its value is fixed once the principle of gauge invariance is specified. 
The apparent arbitrariness instead reflects that the chiral structure of the known quarks and leptons permits an uncountably infinite number of anomaly-free $U(1)$ symmetries, with $\epsilon$ serving to label the distinct theories within this moduli space. 
Crucially, $\epsilon$ is connected to the weak-gravity conjecture~\cite{Arkani-Hamed:2006emk}, which asserts that gravity should be the weakest force. 
Even for extremely small $g_{B-L} < (m_\nu/M_{\rm Pl}) \sim 10^{-30}$, the combination $\epsilon^{-1} g_{B-L}$ can exceed $(m_\nu/M_{\rm Pl})$, ensuring that the enhanced $B-L$ gauge interaction can be stronger than gravity.

We believe the main reason the enhanced $B-L$ symmetry has been overlooked is that it forbids not only the Majorana neutrino mass terms but also the Dirac mass terms. 
Yet the problem (puzzle) for Dirac neutrinos is not how to generate their masses, but explaining why the mass hierarchy between neutrinos and charged fermions is so large. From this prospective, symmetries forbidding Dirac neutrino masses are also motivated. 
Following the Froggatt–Nielsen mechanism~\cite{Froggatt:1978nt}, these masses can arise naturally from the induced sub-eV scale vacuum expectation values; see Ref.~\cite{Ma:2015raa, Ma:2015mjd, Bonilla:2016zef, CentellesChulia:2016rms,  Bonilla:2016diq, Ma:2016mwh, CentellesChulia:2018gwr, Ma:2026tyk} for representative examples, although this approach requires additional model building.

Instead, we associate the spontaneous breaking of the enhanced $B-L$ gauge symmetry to the sub-eV scale neutrino condensate models~\cite{Dvali:2016uhn, Addazi:2016oob, Dvali:2017mpy, deLima:2022dht} (see Refs.~\cite{Barenboim:2008ds, Azam:2010kw, Barenboim:2010db, Barenboim:2019fmj} for discussions on high condensation scales). 
Here, the symmetry-breaking order parameters do not originate from additional fundamental Higgs fields, but instead emerge from the non-perturbative gravitational effects. 
As pointed in Ref.~\cite{Dvali:2017mpy}, no additional assumptions, except for a non-zero topological vacuum susceptibility in pure gravity (in analogy to the $\theta$ term in QCD), are needed. 
The resulting effective neutrino mass term are:
\begin{equation}
    \label{neutrinomass}
    \mathcal{L}~\supset~m^{ij}_{D} \overline{\nu}_{iL}\nu_{iR}, \quad m^{ij}_{D}~\equiv~\frac{1}{\Lambda_G^2}\langle \nu_{iL}\overline{\nu}_{jR}\rangle. 
\end{equation}
Here, $m^{ij}_{D}$ originates from the neutrino condensate order parameter $\langle \nu_{iL}\overline{\nu}_{jR}\rangle$, which is in analogy to $\langle q_L\overline{q}_R \rangle$ of QCD. The difference is all neutrinos acquire non-zero masses irrespective of whether they are confined into bound states or not.
The matrix $m^{ij}_{D}$ can be further diagonalized by a bi-unitary transformation but in general can not be simultaneously diagonalized together with the mass matrix for charged leptons, rendering neutrino mixing angles physical.

The order parameters $\langle \nu_{iL}\overline{\nu}_{jR}\rangle$ break the chiral flavor symmetry of the three generations of massless $\nu_L$ and $\nu_R$ down to one of its maximal anomaly-free subgroups.
Consequently, the enhanced $B-L$ symmetry is spontaneously broken 
and the corresponding gauge boson $A'$ acquires a mass. If $\epsilon^{-1}g_{B-L}\sim \mathcal{O}(1)$, its mass is bounded from below by the neutrino condensate scale, which always lies above the sub-eV range. Heavier $A'$ masses can be generated through to the Stueckelberg mechanism~\cite{Stueckelberg:1938hvi} or additional symmetry breaking order parameters.

\textit{Neutrino Decay and Scattering:}~In the limit $g_{B-L}\to 0$ and $\epsilon\to 0$, with $\epsilon^{-1}g_{B-L}$ finite, $A'$ only couples to two of the three $\nu_R$. Without further extensions, interacts with the other fermions are possible only through the neutrino mass portal which connects $\nu_R$ to $\nu_L$. 
If originated from sub-eV scale order parameters, the neutrino mass $m_{\nu}$ is temperature dependent.
In general, $m_{\nu}$ vanishes above a critical temperature $T_c\sim m_{\nu}$ and becomes non-zero only after a phase transition. If the phase transition is supercooled, $T_c$ can be lower than $m_{\nu}$ and these two scales are independent parameters~\cite{Lorenz:2018fzb}.
In case $T_c<0.1~\text{eV}$ (or equivalently $T_c<10^3~\text{K}$), the early universe observables reduces to those of the standard cosmology with massless $\nu_L$, thereby invaliding the associated constraints, especially the stringent upper bounds on the sum of neutrino masses~\cite{Dvali:2016uhn, Koksbang:2017rux, Lorenz:2018fzb}.

$A'$ is phenomenologically relevant only when $T<T_c$, but the signals are in general suppressed by a chirality flipping factor $m_{\nu}/E$, where $E$ denotes the characteristic energy scale. 
The phenomenological effects of $A'$ become significant only at sufficiently low energies. 
Neutrino decay $\nu_i\to\nu_j A'$ provides a golden channel, in which $E\simeq m_{\nu_i}$ and the chirality flipping suppression is therefore eliminated.

One may expect neutrino decay depends on unknown mixing angles. 
However, following the same logic discussed above on $\epsilon$, 
the information of lepton flavor for $\nu_{R}$ is contained within the principle of gauge invariance.
Through this work, we \textit{define} $(\nu_{eR}, \nu_{\mu R}, \nu_{\tau R})$ for the enhanced $B-L$ symmetry such that in the basis where $m^{ij}_{D}$ --- fixed by non-perturbative gravity --- is diagonal, the $A'$ interaction takes the form: 
\begin{equation}
\label{massbasis}
    \mathcal{L}~\supset~\frac{g_{B-L}}{\sqrt{2}\epsilon}A_{\mu}'
    \left(\overline{\nu}_{3}\gamma^{\mu}P_R \nu_{2}+\overline{\nu}_{2}\gamma^{\mu}P_R \nu_{1}\right)+\text{h.c.},
\end{equation}
If nature does not gauge $U(1)_{B-L}$ in this direction, the flavor structure for neutrino decay changes but stays non-ambiguous as long as the definition of $B-L$ is specified.
Nevertheless, we emphasize that in the flavor alignment limit, where the flavor and mass eigenstates of $\nu_R$ are approximately the same, neutrino decay will be suppressed.

By analogy with the top quark decay $t_L\to b_L W^+$~\cite{Bigi:1986jk}, the $\nu_i\to \nu_j A'$ decay width is:
\begin{equation}
\label{decayR}
    \Gamma^i~=~\epsilon^{-2}g_{B-L}^2\frac{m_{\nu_i}}{64\pi^2}\left(2+\frac{m_{\nu_i}^2}{m_{A'}^2}\right) \left(1-\frac{m_{A'}^2}{m_{\nu_i}^2}\right)^2. 
\end{equation}
Here, we take $m_{\nu_i}\gg m_{\nu_j}$ and for completeness, we present the general expression in the Supplementary Material.
Similar to the top quark decay, the neutrino dominantly decays to the longitudinally polarized $A'$ when $m_{A'}\ll m_{\nu_i}$.
In this limit, $\Gamma^i$ is equal to the $\nu_i\to \nu_j \phi_M$ decay width, where $\phi_M$ is the Majoron of the corresponding gaugeless theory. 
Given the $m_{A'}>m_{\nu_i}$ regime is kinetically forbidden, gauging the enhanced $B-L$ symmetry can not significantly enhance the neutrino decay amplitude. 
Nevertheless, as long as $A'$ is lighter than the heaviest neutrino, the neutrino decay always provides a robust constraint on the enhanced gauge coupling strength $\epsilon^{-1}g_{B-L}$.

A more direct probe for $A'$ is the mediated interaction between neutrinos and charged fermions. In the limit $g_{B-L}\to 0$, additional portals are needed. 
In general, a kinetic mixing term between $A'$ and the visible photon $A$ is allowed:
\begin{equation}
    \label{mixing}
    \mathcal{L}~\supset~\frac{1}{2}\chi F'_{\mu\nu}F^{\mu\nu}. 
\end{equation}
$A'$ can thus also couple with charged particles.
Astrophysics puts stringent constraints on the value of $\chi$; see, e.g. Refs.~\cite{Davidson:2000hf, Hardy:2016kme, Caputo:2021eaa, Li:2023vpv}. 
For $\epsilon^{-1}g_{B-L} \sim \mathcal{O}(1)$, both $A'$ bremsstrahlung and $A'$ mediated right-handed neutrino production can carry energy away during stellar evolution~\cite{Harnik:2012ni, Heeck:2014zfa, Jeong:2015bbi}. Constraints from solar and red-giants cooling then require $\chi\lesssim 10^{-14}$~\cite{Harnik:2012ni}. Hence, neglecting this kinetic mixing term is a good approximation. 
We notice the stringent astrophysics bounds on $\chi$ can, in principle, be evaded similar to the cosmological constraint on $m_{\nu}$. The kinetic mixing $\chi$ can also originate from a symmetry-breaking order parameter and vanish above the critical temperature $T_c$. 
This order parameter may emerge from non-perturbative gravity as both $A'$ and $A$ are contained in the low-energy spectrum and couple directly to gravity. 
We emphasize the distinction: while $m_{\nu}$ does not require more assumptions beyond the $\theta$ term of gravity, the temperature dependent $\chi$ should be understood as an additional hypothesis.

For $T<T_c$, $A'$ mediates $\nu$--$e$ elastic scattering and CE$\nu$NS. In the $m_{A'}\to 0$ limit, the differential cross section $\nu-e$ elastic scattering reads in our natation~\cite{Harnik:2012ni, Bilmis:2015lja, Cerdeno:2016sfi}:
\begin{equation}
\label{scattering}
\begin{aligned}
    \frac{d \sigma_{\nu e} }{d E_r}~=~\epsilon^{-2}g_{B-L}^2\alpha_{EM} \frac{\chi^2 P_{L\to R}}{m_e E_r^2} \left( 2-\frac{2E_r}{E_{\nu}}+\frac{E_r^2}{E_{\nu}^2}\right),
\end{aligned}
\end{equation}
where $E_{\nu}$ is the neutrino energy, and $E_{r}$ is the electron recoil energy. 
We use $P_{L\to R}$ to represent the possibility of observing $\nu_{\mu R}$ or $\nu_{\tau R}$ in the incoming neutrino flux.
$P_{L\to R}$ can directly arise from non-vanishing neutrino masses, in which case it scales as $(m_{\nu}/E_{\nu})^2$ and leads to a suppression. 
In slightly extended scenarios, enhancements are possible.
For instance, if the incoming neutrinos are from the Sun, some of them can lose their chiral origin because solar magnetic fields can induce the neutrino chirality flipping through their magnetic moments;
see, e.g., Ref.~\cite{Barranco:2014cda, Barranco:2017zeq, Joshi:2019dcj} for further analysis. 
Alternatively, if $\nu_R$ acquire tiny Majorana masses from additional order parameters, they become sterile neutrinos which maximally mix with $\nu_L$ while differing in masses, modifying the standard oscillation picture. 
For solar neutrinos, a splitting in the mass square as small as $10^{-10}~\text{eV}^2$ is sufficient to convert a significant fraction of solar neutrinos into $\nu_R$~\cite{Harnik:2012ni}.

The out-going neutrinos from the $A'$ mediated scattering are right-handed polarized.  
Eq~(\ref{scattering}) contains no interference terms with electroweak interactions.
Since the cross section is enhanced at low $E_r$, experiments with lower recoil energy thresholds can provide stronger sensitivities. 
In particular, the proposed superfluid $^4\text{He}$~\cite{vonKrosigk:2022vnf, SPICE:2023tru, DELight:2024bgv} or superconductor~\cite{Hochberg:2015pha, Hochberg:2015fth} based detectors feature the thresholds at eV or lower scales, and are ideal for the ultra-light $A'$~\cite{Alexander:2016aln}.

For light mediator $A'$, although  $\nu-e$ elastic scattering currently provides a stronger constraint on its coupling to the neutrinos and charged fermions, however, as shown Ref.~\cite{Dent:2025drd}, the future low-threshold detectors for CE$\nu$NS can yield comparable bounds. CE$\nu$NS provides a complementary probe, because comparing the two cross sections allows distinguishing our $A'$-visible photon mixing scenario and the other generic models. Moreover, despite the solar neutrino flux is in general larger~\cite{Vitagliano:2019yzm}, the reactor and accelerator neutrino experiments are also important, because their statistics could also be comparable for shorter and shorter distances between the source and detector. This spatial dependence helps to distinguish whether the observed scattering is induced by neutrinos or light dark matter.

\begin{figure*}[t!]
    \centering
    \includegraphics[width=0.9\linewidth]{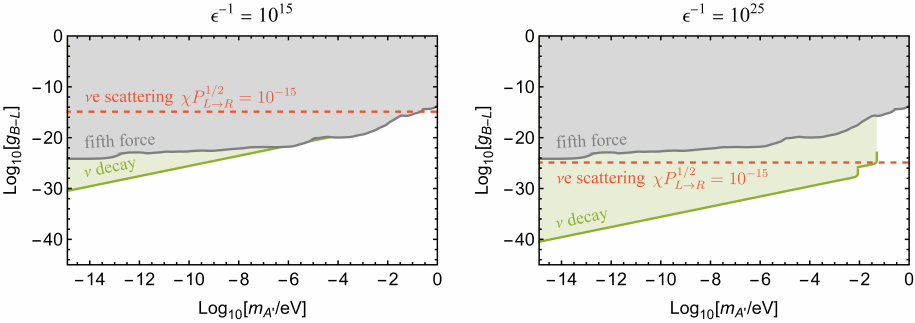}
    \caption{Constraints on the $B-L$ gauge coupling strength $g_{B-L}$, enhanced by $\epsilon^{-1}=10^{15}$ (left panel) and $\epsilon^{-1}=10^{30}$ (right panel), as a function of $m_{A'}$. The gray regions are excluded by the high-precision tests on the fifth force. The green region denotes bounds from the neutrino lifetime. The red dashed line shows the current limit from $\nu-e$ elastic scattering, evaluated with the benchmark value $\chi P_{L\to R}^{1/2}=10^{-15}$.}
    \label{BLbound}
\end{figure*}

\textit{Exclusion Limits}~In Figure~\ref{BLbound}, we show the constraints on the gauge coupling $g_{B-L}$ as a function of $m_{A'}$. We compare $\epsilon^{-1}=10^{15}$ and $\epsilon^{-1}=10^{30}$ to illustrate the $\epsilon^{-1}$ enhancement.

Many astrophysical and terrestrial experiments can constrain the neutrino lifetime; for instance, see Refs~\cite{Ando:2003ie, Fogli:2004gy, Bustamante:2016ciw, Denton:2018aml, Abdullahi:2020rge,  Martinez-Mirave:2024hfd, Valera:2024buc} and Refs~\cite{Gonzalez-Garcia:2008mgl, Gomes:2014yua, Pagliaroli:2016zab, Gago:2017zzy, Choubey:2018cfz, Huang:2018nxj, Porto-Silva:2020gma, Picoreti:2021yct, Ivanez-Ballesteros:2023lqa, Dey:2024nzm} respectively. 
We follow Ref.~\cite{Funcke:2019grs} and take the bound for Dirac neutrinos with normal non-degenerate mass ordering:
\begin{equation}
\label{lifetime2}
    \frac{\tau_3}{m_{\nu_3}}~>~2\times 10^{-10}~\frac{\text{s}}{\text{eV}}, \quad \frac{\tau_2}{m_{\nu 2}}~>~1\times 10^{-3}~\frac{\text{s}}{\text{eV}}, 
\end{equation}
which comes from the analysis on long-baseline and solar neutrino data, respectively. 
Given the associated uncertainties, we exclude the constraints from long-traveling astrophysical neutrinos, such as those observed by IceCube~\cite{IceCube:2015rro, Abbasi:2025fjc} and those from SN1987A~\cite{Kamiokande-II:1987idp, Bionta:1987qt, Alekseev:1988gp}. 
We note, however, that if these uncertainties are brought under control, measurement of astrophysical neutrino flavor ratios at IceCube can yield dramatically improved bounds of $\tau/m_{\nu_i}\gtrsim \mathcal{O}(10)~(\text{s}/\text{eV})$ for all neutrino mass eigenstates~\cite{Bustamante:2016ciw, Valera:2024buc}. 
In Figure.~\ref{BLbound}, the region excluded by neutrino lifetime is indicated in green. 
The neutrino masses hierarchy is assumed maximal; we checked that the the generic non-degenerate mass spectra only lead to minor changes in this logarithmic plot.

Various neutrino and dark matter detection experiments can constrain the $A'$ coupling via $\nu-e$ elastic scattering and CE$\nu$NS; for instance, see recent analysis in Ref.~\cite{Khan:2022bel, Khan:2019cvi,  Khan:2020csx, Dev:2021xzd, CONUS:2021dwh, Chakraborty:2021apc, A:2022acy, DeRomeri:2022twg, DeRomeri:2024iaw, Demirci:2025qdp}.
We fix $\chi P_{L\to R}^{1/2}=10^{-15}$ as a benchmark and show the corresponding bound for $g_{B-L}$ with the red dashed line.
This limit is extracted from the universal tensor model analyzed in Ref.~\cite{Demirci:2025qdp}.


Without the $\epsilon^{-1}$ enhancement, $A'$ directly couples to the neutrons with non-vanishing $g_{B-L}$\footnote{$A'$ can also couple to the neutrons through mixing with the $Z$ boson~\cite{Chauhan:2020mgv, Chauhan:2022iuh}. With the $\epsilon^{-1}$ enhancement, the induced coupling strength scales as $\epsilon^{-1}g_{B-L}G_Fm_{D}^2/(16\pi^2)\simeq \epsilon^{-1}g_{B-L}\times 10^{-28}$.} and thus violates the weak equivalence principle and the inverse-square law of gravity.
For comparison, we show the exclusion regions from the high-precision tests on gravity in gray. 
The bounds summarized in Ref.~\cite{Heeck:2014zfa} remain mostly unchanged and we adopt most of them directly. 
We update the constraints in the mass ranges $m_{A'}\lesssim 10^{-13}$ eV and $m_{A'}\gtrsim 10^{-2}$ eV, which is improved by roughly one order of magnitude thanks to the updated results from the MICROSCOPE satellite~\cite{MICROSCOPE:2019jix, MICROSCOPE:2022doy} (interpreted by Ref.~\cite{Amaral:2024tjg}) and the IUPUI group~\cite{Chen:2014oda}, respectively. 
Comparing with these baryon-based high-precision tests, 
the neutrino experiments can provide stronger constraints on $g_{B-L}$ when $\epsilon$ is sufficiently small, making them important complements probes.

\textit{Conclusion and Outlook:}~The absence of bare neutrino mass terms allows the $U(1)_{B-L}$ charge assignments to be reconsidered beyond their conventional form. Relaxing the equality between $\nu_R$ and $\ell_L$ charges permits anomaly-free solutions in which two generations of $\nu_R$ carry enhanced $B-L$ charges, while all other fermions retain their canonical values. This opens a previously unexplored regime in which the $B-L$ interaction can be strong for neutrinos yet remain extremely weak for baryons and charged leptons. 
Consequently, the $B-L$ gauge coupling, which emerges as a fundamental parameter, can be potentially as large as $\mathcal{O}(1)$, subject only to phenomenological constraints.

Probing the enhanced $B-L$ gauge symmetry is essential for a complete understanding of the Standard Model.
We argue that an immediate experimental priority is to bound the enhanced gauge coupling strength $\epsilon^{-1}g_{B-L}$, in particular to determine whether it is smaller than $\mathcal{O}(1)$. 
From a purely phenomenological perspective, this framework is also of particular interest because it evades the cosmological and astrophysical constraints, serving as an important benchmark for ultralight new physics in general.

\hspace{10pt}

\begin{acknowledgments}
We would like to thank Felix Kahlhoefer, Markus Mosbech, Ulrich Nierste, and Liangliang Su for valuable discussions. 
This research was supported by the Deutsche Forschungsgemeinschaft (DFG, German Research Foundation) under grant 396021762 - TRR 257. X.G. also acknowledges the support by the Doctoral School “Karlsruhe School of Elementary and Astroparticle Physics: Science and Technology.”
\end{acknowledgments}

\appendix
\section*{Supplementary Material}
\noindent
The complete $\nu_i\to \nu_j A'$ decay rate reads:
\begin{equation}
\label{decayR2}
\begin{aligned}
    \Gamma^i~=&~\epsilon^{-2}g_{B-L}^2\frac{m_{\nu_i}}{64\pi^2 } \frac{\lambda^{1/2}(m_{\nu_i}^2, m_{\nu_j}^2,m_{A'}^2)}{m_{A'}^2}\\
    &\times\left( \left(1-\frac{m_{\nu_j}^2}{m_{\nu_i}^2}\right)^2
    +\left(1+\frac{m_{\nu_j}^2}{m_{\nu_i}^2}\right)\frac{m_{A'}^2}{m_{\nu_i}^2}-2\frac{m_{A'}^4}{m_{\nu_i}^4}\right),
\end{aligned}
\end{equation}
where $\lambda(\alpha,\beta,\gamma)~=~\alpha^2+\beta^2+\gamma^2-2\alpha\beta-2\alpha\gamma-2\beta\gamma.$ Eq~(\ref{decayR2}) reduces to Eq~(\ref{decayR}) when $m_{\nu_j}=0$. 
A non-negligible $m_{\nu_j}$ can not change $\Gamma^i$ by orders of magnitude, unless the neutrino masses are in the degenerate limit, in which case $\Gamma^i$ is kinetically suppressed.
Since the KATRIN experiment constrains the absolute neutrino masses~\cite{KATRIN:2024cdt}, this suppression can not be arbitrarily large.


\bibliography{refs}

\end{document}